\documentclass[aps,prd,12pt,preprint,tightenlines,superscriptaddress,
amsfonts,amssymb,amsmath,byrevtex,showpacs]{revtex4}

\begin{document}
%tightenlines - single space manuscript
%eqsecnum - number equations by section
\newcommand{\dR}{\mathbb R}
\newcommand{\dC}{\mathbb C}
\newcommand{\dS}{\mathbb S}
\newcommand{\dZ}{\mathbb Z}
\newcommand{\dN}{\mathbb N}
\newcommand{\dQ}{\mathbb Q}
\newcommand{\id}{\mathbb I}
\newcommand{\ep}{\epsilon}
\newcommand{\dV}{\mathbb V}
\newcommand{\dH}{\mathbb H}
\newcommand{\dM}{\mathbb M}

\title{On three quantization  methods \\for particle on
hyperboloid}
\author{Jean-Pierre Gazeau }
\affiliation{Boite 7020, APC, CNRS
UMR 7164,
Universit\'{e} Paris 7
Denis-Diderot, 75251 Paris Cedex 05,
France; e-mail: gazeau@ccr.jussieu.fr}
\author{Marc Lachi\`{e}ze-Rey}
\affiliation{APC, CNRS UMR 7164, Service d'Astrophysique C.E. Saclay
91191
Gif sur Yvette cedex, France; e-mail: marclr@cea.fr}
\author{W\l odzimierz Piechocki}
\affiliation{Theory Division, So\l tan Institute for Nuclear Studies,\\
Ho\.{z}a 69, 00-681 Warszawa, Poland; e-mail: piech@fuw.edu.pl}

\date{\today}

\begin{abstract}
We compare the  respective efficiencies   of three  quantization
methods (group theoretical, coherent state and geometric) by
quantizing the dynamics of a free massive particle  in
two-dimensional de Sitter space. For each case we consider the
realization of the principal series representation of $SO_0(1,2) $
group and its two-fold covering $SU(1,1)$. We demonstrate that
standard technique  for finding an irreducible representation within
the geometric quantization scheme fails. For consistency we recall
our earlier results concerning the other two methods, make  some
improvements and generalizations.
    \end{abstract}
\maketitle

\section{Introduction}

The aim of this work is to compare  three  quantization methods,
namely group theoretical, coherent state and geometric, by quantizing
the dynamics of a free massive particle  on two-dimensional one-sheet
hyperboloid embedded in three dimensional Minkowski space.

The problem of a particle in 1+1 de Sitter space has already been
considered in  various contexts \cite{TOP,PRV,debire,WP1,WP2}. In
Ref. \onlinecite{GPW} we have carried out the coherent state
quantization, but without taking into account the time-reversal
invariance of the system. The present work feels the gap and presents
generalization of the discussion given in Ref. \onlinecite{GPW}.

In Sec. II we specify the canonical structure of our physical system:
the physical phase space has the topology of a one-sheet hyperboloid,
the basic observables satisfy the  $sl(2,\dR)$ algebra, and the
symmetry group is $SO_0(1,2)$ or $SU(1,1)$ (taking into account the
time-reversal invariance of the dynamics). We carry out the group
theoretical quantization in Sec. III. We get the representation of
$sl(2,\dR)$ algebra by applying the Stone theorem to the principal
series representation of $SU(1,1)$ group. For comparison, we recall
the derivation of the representation of $sl(2,\dR)$ by using the
Schr\"{o}dinger representation for the canonical variables and
applying the symmetrization prescription for basic observables. We
also make comment on the importance of the canonical variables
topology for quantization procedure. Sec. IV concerns coherent state
quantization. We recall basic notions and steps of the method, choose
coherent states suitable to our physical system, carry out mapping of
classical observables into quantum operators, and identify obtained
representation with the Bargmann principal series representation of
$SU(1,1)$. We also extend the analysis in order to include the
time-reversal invariance of the system. The geometric quantization of
our classical system is presented in Sec. V in its prequantization
step. As is known, this method leads usually to reducible
representations. We demonstrate that the  standard technique for
solving the problem does not lead to irreducibility. Our result may
be known (in the context of  principal series representations) to the
community of experts in geometric quantization. However, we have not
found the proof. Our results seem to feel the gap. Each section
includes the discussion of a given quantization method. General
discussion is presented in Sec. VI where we indicate merits and
demerits of the quantization schemes under consideration, and
   we compare their   efficiencies.

We believe that a  generalization of  our results may allow to
quantize the  dynamics of the particle  in realistic four dimensional
curved spacetimes.  Also,  the analysis presented here  may be useful
for the  choice of a quantization scheme in quantum cosmology. For
instance, to address the problem of existence of a big-bounce or
big-crunch/big-bang transition, during the evolution of the universe.
One may examine that issue by considering  propagation of p-brane
states across an orbifold (e.g. Misner's type) singularity (see, Ref.
\onlinecite{bcb} and references therein). Non-perturbative methods of
quantization, discussed in our paper, may overcome technical problems
encountered by perturbative approaches.

\section{Phase space}
\subsection{Particle in de Sitter spacetime}

The two-dimensional de Sitter spacetime $\dV$ has the topology
$\dR^1\times\dS^1$. It  may be visualized as a one-sheet hyperboloid
$\dH_{r_0} $ embedded in 3-dimensional Minkowski space $\dM $, i.e.
\begin{equation}\label{hipp}
\dH_{r_0}:=\{(y^0,y^1,y^2)\in \dM~|~ (y^2)^2+(y^1)^2 - (y^0)^2
=r_0^2,~r_0 >0\},
\end{equation}
where $r_0$ is the parameter of the one-sheet hyperboloid $\dH_{r_0}
$.
The induced metric,
$g_{\mu\nu}~(\mu ,\nu =0,1)$, on $\dH_{r_0} $ is the de Sitter
metric.

An action integral, $\mathcal{A}$, describing a free relativistic
particle of mass $~m_0 >0$  in gravitational field  $g_{\mu \nu}$ is
proportional to the length of a particle world-line:
\begin{equation}\label{lag}
\mathcal{A}=\int_{\tau_1}^{\tau_2}~L(\tau)~d\tau,~~~~~L(\tau) :=-m_0
\sqrt{g_{\mu\nu}~\dot{x}^\mu ~ \dot{x}^\nu} ,
\end{equation}
where $\tau$ is an evolution parameter, $x^\mu$ are intrinsic
coordinates and   $\dot{x}^\mu := dx^\mu/d\tau $. It is assumed that
$ \dot{x}^0>0 $, i.e., $x^0$ has interpretation of time monotonically
increasing with $\tau$.

The action (2) is invariant under the reparametrization $
\tau\rightarrow f(\tau)$ of the world-line (where $f$ is an arbitrary
function of $\tau$). This gauge symmetry leads to the constraint
\begin{equation}\label{con1}
       G:= g^{\mu\nu}p_\mu p_\nu - m_0^2=0,
\end{equation}
where $g^{\mu\nu}$ is the inverse of $g_{\mu\nu}$ and the $p_\mu :=
\partial L/\partial\dot{x}^\mu $ are
canonical momenta.

Since a test particle does not modify spacetime, the local symmetries
of  the system are described by the algebra of all Killing vector
fields.
It is  known  that a Killing vector field $Y$ defines  a dynamical
integral $D$ of a test particle moving along a
geodesic by
\begin{equation}\label{dyn}
D=p_\mu Y^\mu,~~~~~\mu= 0,1 ,
\end{equation}
where $Y^\mu$ are components of $Y$.

To be more specific we parametrize the hyperboloid (\ref{hipp}) as
follows \cite{WP1}
\begin{equation}\label{par}
y^0=-\frac{r_0\cos \rho /r_0}{\sin \rho /r_0},~~~~y^1=\frac{r_0\cos
\theta /r_0} {\sin \rho /r_0},~~~~y^2=\frac{r_0\sin \theta /r_0}{\sin
\rho /r_0},
\end{equation}
where $0<\rho < \pi r_0 $ and $0\leq \theta < 2\pi r_0$.
The metric tensor  on
$\dH_{r_0}$ reads
\begin{equation}\label{met}
ds^2 = (d\rho^2 - d\theta^2)\sin^{-2}(\rho/r_0).
\end{equation}
Thus the constraint (\ref{con1}) has the form
\begin{equation}\label{con2}
G= (p_\rho^2 - p_\theta^2)\sin^2(\rho/r_0)-m_0^2 =0,
\end{equation}
where $p_\rho :=\partial L/\partial\dot{\rho}$ and $p_\theta
:=\partial L/\partial\dot{\theta}$ are canonical momenta.

\subsection{Dynamical integrals from Killing vectors}

The three  Killing vector fields $Y_a ~(a=0,1,2)$ of $\dV$ correspond
   \cite{WP1}    to the   generators  of the proper orthochronous Lorentz
group $SO_0(1,2)$. The infinitesimal transformations, in
parametrization (\ref{par}), read
\begin{equation}\label{t1}
(\rho,~\theta)\longrightarrow(\rho,~\theta+a_0 r_0),
\end{equation}
\begin{equation}\label{t2}
(\rho,~\theta)\longrightarrow(\rho-a_1 r_0 \sin \rho /r_0~\sin \theta
/r_0,~\theta+a_1 r_0\cos \rho /r_0~\cos \theta /r_0),
\end{equation}
\begin{equation}\label{t3}
(\rho,~\theta)\longrightarrow(\rho+a_2 r_0\sin \rho /r_0~\cos \theta
/r_0,~\theta+a_2 r_0\cos \rho /r_0~\sin \theta /r_0),
\end{equation}
where $(a_0,a_1, a_2) \in \dR^3 $ are infinitesimal parameters. The
transformation (\ref{t1}) corresponds to the infinitesimal spatial de
Sitter translations, whereas (\ref{t2}) and (\ref{t3}) define  two
infinitesimal ``boosts" in the embedding spacetime. One of them can
be interpreted as Lorentz boost in de Sitter spacetime;  the other
describes
      de Sitter `time' translation.

The three  corresponding dynamical integrals  (\ref{dyn})  read
\begin{equation}\label{d0}
J_0=p_{\theta}~r_0,
\end{equation}
\begin{equation}\label{d1}
J_1=-p_\rho ~r_0\sin \rho /r_0~\sin \theta /r_0 + p_\theta ~r_0\cos
\rho /r_0~\cos \theta /r_0,
\end{equation}
\begin{equation}\label{d2}
J_2= p_\rho ~r_0\sin \rho /r_0~\cos \theta /r_0 + p_\theta ~r_0\cos
\rho /r_0~\sin \theta /r_0.
\end{equation}

\noindent Making use of  (\ref{d0})-(\ref{d2}) one may rewrite the
constraint (\ref{con2}) as
\begin{equation}\label{hip}
J_2^2 + J_1^2 -J_0^2 =\kappa^2 ,~~~~\kappa:= m_0 r_0,
\end{equation}
where $m_0$ is  the particle mass  and $r_0$ the radius of the
spacetime hyperboloid.

Eqs. (5) and (11)-(13) lead to the algebraic equations\cite{WP1}
   \begin{equation}\label{tra}
J_a y^a =0,~~~~~J_2 y^1 - J_1 y^2 =r_0^2 p_\rho.
\end{equation}
For our system, the physical phase space $\Gamma$ is defined (Ref.
\onlinecite{WP1}) as  the space of all particle geodesics consistent
with the constraint (\ref{con2}). Since each triple $(J_0,J_1,J_2)$
satisfying (\ref{hip}) defines uniquely a particle geodesic (by
solution of (\ref{tra})), the one-sheet hyperboloid (\ref{hip})
represents $\Gamma$. The dynamical integrals $J_a ~ (a=0,1,2) $ are
the basic observables of our system.

\subsection{Gauge system}

The  free particle in curved spacetime, defined by the
action (\ref{lag}), may be treated as a gauge system, with
reparametrization invariance as
gauge symmetry. It is a
characteristic feature of such a system that the Hamiltonian
corresponding to the Lagrangian (\ref{lag}) identically vanishes.
      The general treatment of such gauge system within the constrained
Hamiltonian formalism and the reduction scheme to gauge invariant
variables has been presented elsewhere (see Ref. \onlinecite{GJWP}).
Here we only outline the method leading to the phase space with
independent canonical variables.

For a spacetime dimension $N$, the Hamiltonian formulation of a
theory with gauge invariant Lagrangian \cite{LDF,MHT} leads to an
\emph{extended  phase space} $\Gamma_e$ of dimension $2N$, with  $M$
first-class constraints ($N=2$, $M=1$ in our case). $\Gamma_e$  has a
natural parametrization by $(\rho , \theta ,p_\rho, p_\theta)$. The
constraint surface $\Gamma_c \subset \Gamma_e$, defined by
(\ref{con2}), plays a special role in the formalism.  This  type of
system may have up to $2N-2M$ ($2$ in our case) gauge invariant
functionally independent variables on $\Gamma_c$. Those  may be used
to parametrize the physical phase space and gauge invariant
observables \cite{LDF}. It is known \cite{GJWP} that the dynamical
integrals may be used to represent such variables.

The observables $J_0 , J_1 $ and $J_2$ are gauge invariant (their
Poisson bracket with $G$  vanish) and any two of them are
functionally independent on $\Gamma_c$ due to (\ref{hip}).    There
exists  a general method for finding the corresponding canonical
variables, although it is quite involved \cite{GJWP}. In what follows
we recall the simple method used in Ref. \onlinecite{WP1}. It
consists of three steps:

\noindent First, we identify the algebra of the   observables
on   $\Gamma_e$.
The canonical coordinates above  define  the Poisson bracket on
$\Gamma_e$,
\begin{equation}\label{surcon}
\{\cdot,\cdot\}:=\frac{\partial \cdot}{\partial p_\rho}\frac{\partial
\cdot}{\partial \rho} - \frac{\partial \cdot}{\partial
\rho}\frac{\partial \cdot}{\partial p_\rho} + \frac{\partial
\cdot}{\partial p_\theta}\frac{\partial \cdot}{\partial \theta} -
\frac{\partial \cdot}{\partial \theta}\frac{\partial \cdot}{\partial
p_\theta}  .
\end{equation}
Direct calculations lead to
\begin{equation}\label{clcom}
\{J_0,J_1\}=-J_2,~~~\{J_0,J_2\}=J_1,~~~\{J_1,J_2\}=J_0 .\end{equation}
   Hence, the three basic observables $J_a$ satisfy the $sl(2,\dR)$
algebra.

\noindent Second, we parametrize   the physical phase space $\Gamma$,
identified with the hyperboloid (\ref{hip}), by two coordinates $J,
\beta$   defined through
\begin{equation}\label{param}
J_0 := J,~~~J_1:=J\cos\beta - \kappa\sin\beta,~~~J_2:=J\sin\beta +
\kappa\cos\beta,~~~0<\kappa <\infty .
\end{equation} Thus, we identify it with
\begin{equation}\label{paraph}
X:=\{x \equiv (J,\beta)~|~J \in \dR, ~ 0\leq \beta
<2\pi\}.\end{equation}

Third, we write a  Poisson bracket in $X$,  with  the
variables $J$ and  $\beta$:
\begin{equation}\label{pois}
\{\cdot,\cdot\}:=\frac{\partial\cdot}{\partial J}\frac{\partial\cdot}
{\partial \beta}-\frac{\partial\cdot}{\partial
\beta}\frac{\partial\cdot}{\partial J}~.
\end{equation}
Since $\{J,\beta\}=1$,   $~J$ and $\beta$ are  canonical variables.
In what follows, we call   $X$, expressed in terms of them,
   the  \emph{canonical phase space}. This specifies
completely  the canonical structure of our system.

\subsection{Symmetries}

The local symmetries of the phase space  coincide with the local
symmetries of the Lagrangian of a free particle in de Sitter's space.

Now, let us discuss the symmetry group of the phase space $\Gamma$.
There are infinitely many Lie groups having $~sl(2,\dR)\sim
su(1,1)\sim so(1,2)$ as their Lie algebras. The common examples are:
\begin{itemize}
    \item
$SO_0(1,2)$, the proper orthochronous Lorentz group;
    \item
$SU(1,1) \sim
SL(2,\dR)$, the two-fold covering of $SO_0(1,2)\sim SU(1,1)/\dZ_2$;
    \item
$~\widetilde{SL(2,\dR)}$, the infinite fold covering of
$SO_0(1,2)\sim \widetilde{SL(2,\dR)}/\dZ $ (the universal covering
group). \end{itemize} Let us give  a  short description of the way in
which the $SU(1,1)$ symmetry appears in this problem, and what are
the precise connections with the $SO_0(1,2)$ and
$\widetilde{SL(2,\dR)}$ (possible) symmetries. The action of
$SU(1,1)$  on the de Sitter spacetime can be understood through the
following ``space-time'' factorization\cite{JGH} of a generic element
$g =
\begin{pmatrix}\label{fact}
a & b \\
\bar{b} & \bar{a}
\end{pmatrix}, \ a,b \in \dC, \ \vert a\vert^2 -
\vert b \vert^2 = 1$ of the group:
\begin{equation}\label{spf}
\begin{split}
     g  &=\underset{j}{\underbrace{\underset{``space"\,
translation}{\underbrace{\begin{pmatrix}
e^{i\theta/2} & 0 \\
0 & e^{-i\theta/2}
\end{pmatrix}}}
\underset{``time"\, translation}{\underbrace{\begin{pmatrix}
\cosh{\frac{\psi}{2}} & \sinh{\frac{\psi}{2}} \\
\sinh{\frac{\psi}{2}} & \cosh{\frac{\psi}{2}}
\end{pmatrix}}}}}
\underset{l}{\underbrace{\underset{Lorentz\,
tranformation}{\underbrace{
\begin{pmatrix}
\cosh{\frac{\varphi}{2}} & i\sinh{\frac{\varphi}{2}} \\
-i\sinh{\frac{\varphi}{2}} & \cosh{\frac{\varphi}{2}}
\end{pmatrix}}}}}\\
& \equiv s(\theta)~ t(\psi) ~l(\varphi)  \equiv j(\theta, \psi)
~l(\varphi), \ \mbox{with} \ 0\leq \theta < 4 \pi,~~~  \varphi, \psi
\in \dR.
\end{split}
\end{equation}
Note that the parameter $\varphi$ stands for the Lorentz rapidity.

The  right coset $\mathcal{J} = SU(1,1)/(\mbox{Lorentz subgroup})  $
is the subset of elements of the $j$ type.
The space-time factorization (\ref{spf}) allows us to make $SU(1,1)$
acts on
$\mathcal{J}$ by left action,
$$SU(1,1) \ni g:\  j \rightarrow j' \equiv g\cdot j,$$ where $j'$
is defined by $ gj = j'l'$. From this, we  easily infer the action of
$SU(1,1)$ on the matrices of
the type $jj^t$:
\begin{equation}\label{ggtact}
g:\  jj^t \rightarrow j'{j'}^t = gjj^tg^t.
\end{equation}

On the other hand, each group element of the type $jj^t$ is in
one-to-one correspondence with a point of the double covering of
$\dH_{r_0} $ defined by  Eq. (\ref{hipp}), through
\begin{equation}\label{matds}
jj^t = \begin{pmatrix}
e^{i\theta}\cosh{\psi} & \sinh{\psi} \\
\sinh{\psi} & e^{-i\theta}\cosh{\psi}
\end{pmatrix} \equiv \begin{pmatrix}
r_0^{-1}y^+ & r_0^{-1}y^0 \\
r_0^{-1}y^0 & r_0^{-1}y^-
\end{pmatrix}, \ \mbox{with} \ y^{\pm}= y^1 \pm i y^2.
\end{equation}
This  provides  global coordinates for this double covering:
\begin{equation}\label{param1}
y^0 = r_0\sinh{\psi}, ~~~ y^1 = r_0\cos{\theta}\cosh{\psi}, ~~~ y^2 =
r_0\sin{\theta}\cosh{\psi}.
\end{equation}
Hence, we easily derive from (\ref{ggtact}) the homomorphism which
sends $SO_0(1,2)$ into $SU(1,1)$.

\subsection{Orbits of the Lie algebra $su(1,1)$}

Let us chose for $su(1,1)$, the Lie algebra of $SU(1,1)$, a basis
$(Y_s, Y_t, Y_l)$ corresponding
to the above {\it space-time-Lorentz } parametrization:
\begin{equation}\label{stl}
Y_s = \frac{1}{2}\begin{pmatrix}
i & 0 \\
0 & -i
\end{pmatrix}, ~~~ Y_t = \frac{1}{2}\begin{pmatrix}
0 & 1 \\
1 & 0
\end{pmatrix}, ~~~ Y_l = \frac{1}{2}\begin{pmatrix}
0 & i \\
-i & 0
\end{pmatrix}.
\end{equation}
These generators  obey the   commutation rules
\begin{equation}\label{alg3}
\lbrack Y_t, Y_l\rbrack = - Y_s, ~~~ \lbrack Y_l, Y_s\rbrack =  Y_t,
~~~ \lbrack Y_s, Y_t\rbrack =  Y_l.
\end{equation}
A generic element $Y$ of  $su(1,1)$ reads:
\begin{equation}\label{ygr}
Y = \xi_s Y_s + \xi_t Y_t + \xi_l Y_l =\begin{pmatrix}
iu & \zeta \\
\bar{\zeta} & -iu
\end{pmatrix} .
\end{equation}

Since $su(1,1)$ is a simple Lie algebra, the bilinear form
$\langle Y_1,Y_2 \rangle \equiv \mbox{Tr}Y_1Y_2 $
is nondegenerate
and allows us to identify $su(1,1)$ and its vector space dual.
Therefore, the classification of the co-adjoint orbits as possible
phase spaces for motions on the de Sitter space amounts to
classify orbits of the adjoint representation:
\begin{equation}
\label{adjrep}
su(1,1) \ni Y \stackrel{g}{\rightarrow} Y' = g Y g^{-1}.
\end{equation}
The trivial one apart, three types of orbits are found. The first family
corresponds to the transport of the
particular element $2 \kappa Y_t$.
The subgroup stabilizing this element under the adjoint action is  the
non-compact $SO(1,1)$ corresponding to  the time-translation subgroup
for the de Sitter  space-time, since the  phase space for a test
particle in de Sitter space-time can be viewed as the group coset
$SU(1,1)/$(``Time-translation subgroup'').  By using Lorentz boosts and
dS space  translations
    only for transporting the latter, we obtain   orbit
generic  elements as
\begin{align}\label{dsorb}
\nonumber Y(J, \beta) &=s(\theta) l(\varphi) \,2\kappa Y_t\;
l(-\varphi) \;s(-\theta) \\
\nonumber &=\kappa\,\begin{pmatrix}
i\sinh{\varphi} & \cosh{\varphi} \;e^{i\theta}\\
\cosh{\varphi} \;e^{-i\theta} & - i\sinh{\varphi}
\end{pmatrix} \\
\nonumber& \equiv \begin{pmatrix}
iJ & p_0 \;e^{i \beta} \\
p_0 \;e^{-i \beta} & -iJ
\end{pmatrix}, \\
&\mathrm{with} ~~~ J = \kappa \sinh{\varphi}, ~~~ p_0 =
\sqrt{\kappa^2 + J^2 }, ~~~  \theta = \beta   +
\arctan{\frac{\kappa}{J}}.
\end{align}

We recognize here the phase space parameters introduced in
(\ref{param}). Note  that  the invariant $\kappa^2 = \det{Y}$ and
(the Minkowskian-like) ``energy'' $p_0$ is also equal to
     $p_0 = \sqrt{J_1^2 + J_2^2}$). This matrix realization of the
phase-space for  massive
particle is very convenient for describing the action of the symmetry
group, since we simply have $Y(J',\beta') = g Y(J,\beta) g^{-1}$.

Apart from continuous transformations, the phase space may be also
invariant under discrete transformations. The most important seems to
be the time-reversal invariance, since our system is not dissipative
one. It has been shown in Ref. \onlinecite{WP2} that taking into
account of this invariance leads to the conclusion that the symmetry
group of the phase space $\Gamma$ of our system must be   $SU(1,1)$.

\section{Group theoretical quantization}

Now we intend to find (essentially) self-adjoint representations of
the $su(1,1)$ algebra, integrable to unitary irreducible
representations (UIR's) of the $SU(1,1)$ group.

The set of unitary irreducible representations of the  group $SU(1,1)
\simeq SL(2,\mathbb{R})$ is well known since the seminal work of
Bargmann \cite{VB}.  It is made up of three series : the principal
series, the complementary series, and the discrete series. Since we
are concerned with the massive case, we only consider the principal
series. We briefly recall here the construction of the principal
series by using the Mackey's method of induced representations
\cite{lip,taka}. We consider the Iwasawa decomposition of $SU(1,1) =
KAN$, where $K\simeq U(1)$ is the maximal compact subgroup, $A \simeq
SO(1,1)$ is the Cartan maximal abelian subgroup,  and $N \simeq
\mathbb{R}$ is nilpotent:
\begin{equation}
\label{iwas} SU(1,1) \ni g = \begin{pmatrix}
e^{i\theta/2} & 0 \\
0 & e^{-i\theta/2}
\end{pmatrix}
\begin{pmatrix}
\cosh{\frac{\psi}{2}} & \sinh{\frac{\psi}{2}} \\
\sinh{\frac{\psi}{2}} & \cosh{\frac{\psi}{2}}
\end{pmatrix}
\begin{pmatrix}
1 + ix & -ix \\
-ix & 1 - ix
\end{pmatrix} \equiv s(\theta) t(\psi) n(x),
\end{equation}
with $ 0 \leq \theta < 4 \pi$, ~~~$\psi, \, x \in \mathbb{R}$. The
subgroup $M \simeq \mathbb{Z}_2$ is the centralizer of $A$ in $K$.
Then, $AN$ is a solvable subgroup and $B = MAN$ is the (minimal)
parabolic subgroup of $SU(1,1)$. In other words, the principal series
is induced by a character of the subgroup $B$, the Lie algebra of
$AN$ being a real polarization for the differential of this
character.
   Let $\sigma$ and $\tau$ be  UIR of $M$ and $A$
respectively. Actually, $\sigma$ reduces here to $\pm I_d$ (``even''
or ``odd'') and $\tau$ is a character of $\mathbb{R}$. The
\textit{principal series} of representations constitute the family of
representations
\begin{equation}
\label{induced}
\pi(\sigma,\tau) = \mathrm{Ind}_B^{SU(1,1)}(\sigma \times \tau).
\end{equation}
In what follows we describe these representations in the so-called
``compact'' realization.\cite{lip} They are labelled  by the
parameter $\chi = (l, \lambda)$.
\begin{itemize}
      \item Here $l = - \frac{1}{2} -  i \rho, \,  \rho \in \dR$, and
$\lambda = 0$ or $\frac{1}{2}$. Parameter $\kappa$ of this paper
is related to $\rho$ by $\kappa = - \rho$ for $\rho < 0$.
      \item  They act in the Hilbert space $L^2(\lbrack 0, 2\pi))=
\left\{f(e^{i \beta})\ | \ \Vert f \Vert^2 = \frac{1}{2\pi}
\int_0^{2 \pi} \vert f(e^{i \beta})\vert^2 \,d\beta < \infty
\right\}$ of the exponential Fourier series.
      \item  The representation operator $T^{\mathrm{ps}}_{\chi}$ is
given
by:
      \begin{equation}\label{reprinc}
T^{\mathrm{ps}}_{\chi}(g)f(e^{i \beta}) = \left(b e^{i \beta}
+ \bar{a} \right)^{l - \lambda} \, \left( \bar{b} e^{-
i \beta} +\alpha \right)^{l  + \lambda} \, f \left( \frac{
a e^{i \beta} + \bar{b}}{ b e^{i \beta} +
\bar{a}}\right).
\end{equation}
\item The representations $T^{\mathrm{ps}}_{(l, \lambda)}$ and
$T^{\mathrm{ps}}_{(-l - 1, \lambda)}$ are unitarily equivalent.
\end{itemize}

Applying the Stone theorem \cite{SMH} to  (\ref{reprinc}) leads to
the following operators corresponding to the generators of $SU(1,1)$
group
\begin{equation}\label{j0}
\hat{J}_0 =\lambda - i \frac{d}{d\beta},~~~\lambda =0,~1/2, ~~~
\mathrm{with} ~~~ T^{\mathrm{ps}}_{\chi}(s(\theta)) \underset{\theta
\to 0}{\sim} \mathbb{I} + \theta i \hat{J}_0 ,
\end{equation}
\begin{equation}\label{j1}
\hat{J}_1 = \cos\beta \hat{J}_0 - (\kappa-i/2) \sin\beta , ~~~
\mathrm{with} ~~~ T^{\mathrm{ps}}_{\chi}(l(\varphi)) \underset{\theta
\to 0}{\sim} \mathbb{I} - \varphi i \hat{J}_1 ,
\end{equation}
\begin{equation}\label{j2}
\hat{J}_2 = \sin\beta \hat{J}_0 + (\kappa -i/2) \cos\beta , ~~~
\mathrm{with} ~~~ T^{\mathrm{ps}}_{\chi}(t(\psi)) \underset{\theta
\to 0}{\sim} \mathbb{I} - \psi i \hat{J}_2 .
\end{equation}

The operators $\hat{J}_a~~(a=0,1,2)$ are essentially self-adjoint on
a suitable common dense subspace $\Omega_\lambda \subset L^2
[0,2\pi]$. These operators cannot be defined on the entire Hilbert
space because they are unbounded.

The quantum Casimir operator corresponding to (\ref{j0})-(\ref{j2})
reads
\begin{equation}\label{hcas}
\hat{C} := \hat{J}_2^2 + \hat{J}_1^2 -\hat{J}_0^2 = (\kappa^2+1/4)
\id .
\end{equation}
Eq. (\ref{hcas}) shows that there exists $\Omega_\lambda $ such that
the representation (\ref{j0})-(\ref{j2}) may be lifted to the  UIR of
the $SU(1,1)$ group. Since $0< \kappa < \infty$ the representation
belongs to the principal series.

For comparison, let us recall  the method of mapping of the basic
observables into quantum operators commonly used by physicists and
applied in Ref. \onlinecite{WP1}:

\noindent First, one maps the canonical observables $J$ and $\beta$
as follows
\begin{equation}\label{can1}
    J \longrightarrow \hat{J}\Psi(\beta):= (\lambda -
i\frac{d}{d\beta}) \Psi (\beta),
\end{equation}
\begin{equation}\label{can2}
    \beta \longrightarrow \hat{\beta}\Psi(\beta):=\beta \Psi(\beta),
\end{equation}
where $\Psi \in L^2[0, 2\pi]$.

    \noindent
Next, one applies the symmetrization prescription to all products
    in (\ref{param}), i.e.
    \begin{equation}\label{AB}
    AB \longrightarrow \frac{1}{2}(\hat{A}\hat{B} + \hat{B}\hat{A}).
    \end{equation}
    As the result one gets the operators $\hat{J}_a~~(a=0,1,2)$ in the
form
    (\ref{j0})-(\ref{j2}).
    It was shown in Ref. \onlinecite{WP1} that the heuristic mapping
(\ref{can1})-(\ref{can2})
    leads to the homomorphism
    \begin{equation}\label{homm}
[\hat{J}_a,\hat{J}_b] =-i\widehat{\{J_a,J_b\}},~~~~a,b=0,1,2
\end{equation}
and leads to the essentialy self-adjoint representation of $sl(2,\dR)$
algebra
on $\Omega_{\lambda}\subset L^2[0,2\pi]$ defined to be
\begin{equation}\label{omeg}
\Omega_{\lambda} := \{\Psi\in L^2 [0,2\pi]~|~\Psi\in
C^\infty[0,2\pi], ~\Psi^{(n)}(0)= e^{i\lambda \pi}\Psi^{(n)}(2\pi),~
n=0,1,2...\}.
\end{equation}
We emphasize here that the problem of quantization of the algebra of
basic observables (\ref{clcom}) does not depend on the problem of
quantization of the canonical algebra
\begin{equation}\label{jb1}
\{J,\beta\} =1,
\end{equation}
in the sense that one can find an (essentially) self-adjoint
representation of (\ref{clcom}) without the requirement that one
should first find an (essentially) self-adjoint representation of
(\ref{jb1}).

The discussion concerning the quantization of the algebra (\ref{jb1})
has a long history (see  Ref. \onlinecite{R1} and references
therein). One knows that the commonly used Schr\"{o}dinger
representation of (\ref{jb1}) defined by
\begin{equation}\label{sch}
J\longrightarrow \hat{J} \Psi (\beta):= - i \frac{d}{d\beta}\Psi
(\beta)~~~~ \beta\longrightarrow \hat{\beta}\Psi (\beta):=\beta \Psi
(\beta), ~~~~\Psi\in \mathcal{H}
\end{equation}
(where $\mathcal{H}$ is a Hilbert space) can be made (essentially)
self-adjoint, if $(J,\beta)\in \dR\times\dR$, i.e. in case both
operators $\hat{J}$ and $\hat{\beta}$ are unbounded, and if the Weyl
relations are satisfied \cite{R1}.  In such a case one can find a
common dense invariant subspace $\Omega\subset L^2(\dR)$ such that
both $\hat{J}$ and $\hat{\beta}$ are essentially self-adjoint on
$\Omega$. Such result is a consequence of the Stone-von Neumann
theorem.

It has been shown (see App.\:B of Ref. \onlinecite{WP3}) that, if
$\beta \in[a,b] \subset \dR$ the self-adjoint representation of
(\ref{jb1}) obtained by making use of (\ref{sch}) does not exist.
This result can be probably extended to the  case of any
representation of (\ref{jb1}), but such that the quantum operator
$\hat{J}$ (or $\hat{\beta}$) is unbounded whereas $\hat{\beta}$ (or
$\hat{J}$) is bounded.  It was proved in Ref. \onlinecite{R2} that in
case both operators $\hat{J}$ and $\hat{\beta}$ are bounded, the
self-adjoint representation of (\ref{jb1}) cannot exist.

The problem discussed above concerns canonical variables with a
trivial topology. As it is known the entire programme connected with
higher dimensional theories rests heavily on canonical variables with
non-trivial topology. The problem of quantizing a particle on
one-sheet hyperboloid, considered in this paper, is a simple example
of the situation when one of the variables has a non-trivial
topology, i.e.  $(J,\beta) \in \dR^1 \times \dS^1 $. In such a case
(\ref{jb1}) should be replaced by
\begin{equation}\label{juu}
\{J,U\}=U,~~~~~U:=e^{i\beta}.
\end{equation}
It is shown in Ref. \onlinecite{WP2} that  (\ref{juu}) may be used to
impose the self-adjointness onto the algebra of basic observables
already at the canonical level. It is clear that the choice of
canonical variables in the form compatible with their topologies has
basic significance (see \cite{Kastrup:2005xb} for more discussion).

\section{Coherent state quantization}
\label{cosqu}

We start  from the canonical phase space $X$ equipped with some
measure $\mu$, e.g. its canonical phase space measure.  Let ${\cal
H}$ be a separable Hilbert space. Suppose there exists a continuous
mapping
\begin{equation}\label{xX}
  X \ni x \longrightarrow  | x \rangle \in {\cal H}
\end{equation}
(in Dirac notations), defining
a family of states $\{ | x \rangle \}_{x\in X}$ obeying the following
two conditions:
\begin{itemize}
\item {\bf Normalisation}
\begin{equation}\langle \, x\, | x \rangle = 1,
\label{norma}
\end{equation}\item {\bf Resolution of the unity in ${\cal H}$}
\begin{equation}\int_X  | x\rangle \langle x \, | \, \nu(dx)=
\id_{{\cal H}},
\label{iden}
\end{equation}where $\nu(dx)$ is another measure on $X$, usually
absolutely continuous with respect to
$\mu(dx)$: there exists a positive measurable $h(x)$ such that $\nu(dx)
= h(x) \mu(dx)$.
\end{itemize}

The resolution of the unity on ${\cal H}$ can alternatively be
understood in terms of the scalar product $\langle  x | x' \rangle$
of two states of the family. Indeed, Eq. (\ref{iden}) implies that to
any vector $| \phi \rangle$ in ${\cal H}$ one can isometrically
associate the function $\phi(x) \equiv \sqrt{h(x)}~\langle x | \phi
\rangle$ in  $L^2(X, \mu)$, and this function obeys $\phi(x) = \int_X
\sqrt{h(x)h(x')} ~\langle x | x' \rangle ~\phi(x')~\, \mu(dx') $.
Hence, ${\cal H}$ is isometric to a reproducing Hilbert space
$\mathcal{K}$, closed subspace of $L^2(X, \mu)$,  with kernel
$K(x,x') = \sqrt{h(x)h(x')} ~\langle x | x' \rangle$.

The quantization of a {\it classical} observable, that is to say of a
function $f(x)$ on
$X$, having specific properties in relationship with the topological
structure allocated to $X$, simply consists in associating to  $f(x)$
the operator
\begin{equation} A(f) \equiv  \int_X  f(x) ~ | x\rangle \langle x| \,~
\nu(dx).
\label{oper}
\end{equation}
In this context, $f(x)$ is said  to be upper (or contravariant
\cite{berez}) symbol of the operator $A(f)$, whereas the mean value
$\langle x| f(x) | x\rangle$ is said lower (or covariant) symbol of
$A(f)$. Of course, such a particular quantization scheme is
intrinsically limited to all those classical observables for which
the expansion (\ref{oper}) is mathematically justified within the
theory of operators in Hilbert spaces ({\it e.g.} weak convergence).

A method of construction \cite{GGH,AEG} of $| x\rangle$ has a wave
packet flavor, in the sense that it is obtained from some
superposition of elements of an orthonormal basis $\{ | n\rangle
\}_{n \in \dN}$ of ${\cal H}$. Suppose that the basis $\{ | n\rangle
\}_{n \in \dN}$ is in one-to-one correspondence with an orthonormal
set $\{ \phi_n(x) \}_{n \in \dN}$ (as elements of  $L^2(X, \mu)$).
Furthermore, and this is a decisive step in the wave packet
construction, we assume that
\begin{equation}{\cal N} (x) \equiv \sum_n \vert \phi_n (x) \vert^2 <
\infty \ \mbox{almost everywhere}.
\label{factor}
\end{equation}
Then, the states
\begin{equation}| x\rangle \equiv \frac{1}{\sqrt{{\cal N} (x)}} \sum_n
\overline{\phi_n (x)}~ | n\rangle, \label{cs}
\end{equation}satisfy both our  requirements (\ref{norma}) and
(\ref{iden}). Indeed, the normalization follows    automatically  from
the orthonormality of the set  $\{| n\rangle\}$  and from the
presence of the normalization factor (\ref{factor}). The resolution
of the unity in ${\cal H}$  also follows from   the orthonormality
of the same set $\{\phi_n (x)\}$, so long as the measure $\nu(dx)$
is related to $\mu(dx)$ by
\begin{equation}
\nu(dx) = {\cal N} (x)\, \mu(dx).
\label{meas}
\end{equation}

\subsection{Choice of coherent states}

Let us recall, modify and extend the quantization scheme presented in
Ref. \onlinecite{GPW}.

\noindent First, we define the measure $\mu$ on $X$ to be the
canonical one
\begin{equation}
\label{muep} \mu (dx):= d\beta \, dJ/\:2\pi.
\end{equation}
Next, we introduce a subsidiary abstract separable Hilbert space
$\mathcal{H}$ with an  orthonormal  basis $\{|m>\}_{m\in \dZ} $, i.e.
\begin{equation} \label{obas}
<m_1|m_2>= \delta_{m_1,m_2},~~~\sum_{m\in \dZ}|m><m|=\id ,
\end{equation}
and  an orthonormal set of vectors $\{\phi_m^\ep\}_{m\in \dZ}$ which
spans the Hilbert space $L^2(X,\mu)$ appropriate to our physical
system. For regularization purposes we introduce here an arbitrary
small real parameter $\ep >0$.

\noindent Being inspired by the choice of the coherent states for the
motion of a particle on a circle \cite{deb,RTR,KKJ,JAG,KR1},  we
define $\phi_m^\ep$'s to be suitably weighted Fourier exponentials
     \begin{equation}
     \label{phimep}
\phi_m^\ep(\beta, J):=
\big(\frac{\ep}{\pi}\big)^{1/4}\exp\big(-\frac{\ep}{2}(J-m)^2\big)
\exp (im\beta),~~~~~~~m\in \dZ .
\end{equation}

\noindent The coherent states $|x,\ep> \in \mathcal{H}$ are defined
as follows
\begin{equation}\label{cosit}
X\ni x \longrightarrow |x,\ep>\equiv |\beta,J,\ep>:=
\frac{1}{\sqrt{\mathcal{N}_\ep (\beta, J)}}\sum_{m\in
\dZ}\overline{\phi_m^\ep (\beta,J)}|m> ,
\end{equation}
where the normalization factor $\mathcal{N}_\ep$ reads
\begin{equation}\label{theta}
\mathcal{N}_\ep(x):= \sum_{m\in \dZ}|\phi_m^\ep |^2
=\big(\frac{\ep}{\pi}\big)^{1/2}\sum_{m\in \dZ}
\exp\left(-\ep(J-m)^2\right)< \infty ,
\end{equation}
and it is proportional to an elliptic theta function.

By construction, the states (\ref{cosit}) are normalized and lead to
the resolution of the identity in $\mathcal{H}$
\begin{equation}
\label{cositres}
\frac{1}{2\pi}\int_{0}^{2\pi}d\beta\int_{-\infty}^{\infty}dJ\,
\mathcal{N}_\ep(\beta, J)|\beta,J,\ep><\beta,J,\ep|=\id .
\end{equation}

\subsection{Mapping of classical observables}

A class of quantum operators is naturally given in ``CS diagonal
representation''   by the mapping
\begin{equation}
\label{dsoper} C^\infty (X,\dR)\ni f \longrightarrow  \hat{f}^\ep
:=A_\ep(f):= \int_X \mu(dx)\mathcal{N}_\ep (x)f(x)|x,\ep><x,\ep| ,
\end{equation}
     where
     \[ \hat{f}^\ep:L^2 (X,\mu)\rightarrow  L^2 (X,\mu) ,\]
for any   classical observable  $f$, i.e.,  a function of
$(\beta, J)$ with reasonable properties.

     For an arbitrary function of $J$ alone, we get the diagonal
operator:
     \begin{align}
     \nonumber A_\ep(f) &= \int_{X} \mu_\ep (dx) \, \mathcal{N}_\ep(J)
f(J) \,  | \beta,J,\ep \rangle \langle  \beta, J, \ep |  \\
&= \sum_{m\in \dZ} \langle f \rangle_{\ep,m}
| m\rangle \langle m|,
\label{f(J)}
\end{align}
where $\langle f \rangle_{\ep,m}$ designates the mean value of $f(J)$
with respect to the Gauss normal distribution
$\sqrt{\frac{\ep}{\pi}}\,e^{-\ep(J-m)^2}$ centred at $m$ and with width
$\sqrt{\frac{2}{\ep}}$.

     For a function of $\beta$ alone, we have \begin{align}
     \nonumber A_\ep(f) &  =  \int_{X} \mu (dx)\,
\mathcal{N}_\ep(J) f(\beta) \,  | \beta, J, \ep \rangle \langle
\beta, \ep |  \\
&= \sum_{m,m'}
e^{-\frac{\epsilon}{4}\,(m-m')^2} \,c_{m-m'}(f)| m\rangle \langle m' |
\label{f(varphi)}
\end{align}
     where $c_m(f) = \frac{1}{2\pi}\int_0^{2\pi} e^{-im\beta}
f(\beta)\, d\beta$ is the $m$th Fourier coefficient of $f$.

\subsection{Homomorphism}
Let  us introduce
three suitably shifted versions of the fundamental classical
observables (\ref{param}), namely:
\begin{equation}
\label{shifobs} J_0^{\lambda} = J_0 + \lambda, ~~~ J_1^{\lambda} =
J_1 + \lambda \cos{\beta}, ~~~ J_2^{\lambda} = J_2 + \lambda
\sin{\beta}.
\end{equation}
Note that they have the same Poisson brackets as $J_0,~J_1$ and $J_2$
(see (\ref{clcom})) :
\begin{equation}\label{clshiftcom}
\{J_0^{\lambda},J_1^{\lambda}\}=-
J_2^{\lambda},~~~\{J_0^{\lambda},J_2^{\lambda}\}=J_1^{\lambda},~~~\{J_1^
{\lambda},J_2^{\lambda}\}=J_0^{\lambda} .
\end{equation}

A straightforward application of (\ref{f(J)}) and (\ref{f(varphi)})
(compare to Ref. \onlinecite{GPW}) yields the quantum counterparts of
(\ref{shifobs}):
\begin{equation}
\label{J0}
\hat{J}_0^{\ep, \lambda} =A_\ep (J_0^{\lambda}) = \sum_{m\in \dZ}
(m+\lambda)~ |m><m|.
\end{equation}

\begin{equation}
\label{J1}
\hat{J}_1^{\ep, \lambda} = A_\ep (J_1^{\lambda})= \frac{1}{2}e^{-\ep
/4}\sum_{m\in \dZ}
\left((m+\frac{1}{2}+ \lambda
+i\kappa)~|m+1><m| + c. c. \right) ,
\end{equation}
\begin{equation}
\label{J2}
\hat{J}_2^{\ep, \lambda} = A_\ep (J_2^{\lambda})= \frac{1}{2i}e^{-\ep
/4}\sum_{m\in \dZ}
     \left((m+\frac{1}{2}+\lambda
+i\kappa)~|m+1><m| - c. c. \right) ,
\end{equation}
where $c. c. $ stands for the complex conjugate of the preceding term.

The commutation relations corresponding to (\ref{clcom}) are
found to be
\begin{equation}\label{qcom1}
[\hat{J}_0^{\ep, \lambda},\hat{J}_1^{\ep, \lambda}]=i\hat{J}_2^{\ep,
\lambda}=-iA_\ep (\{J_0^{\lambda},J_1^{\lambda}\}),
\end{equation}
\begin{equation}\label{qcom2}
[\hat{J}_0^{\ep, \lambda},\hat{J}_2^{\ep, \lambda}]=-i\hat{J}_1^{\ep,
\lambda}=-iA_\ep (\{J_0^{\lambda},J_2^{\lambda}\}),
\end{equation}
\begin{equation} \label{qcom3}
[\hat{J}_1^{\ep, \lambda},\hat{J}_2^{\ep, \lambda}]=-i e^{-\ep
/2}\hat{J}_0^{\ep, \lambda} =e^{-\ep /2}
\left( -iA_\ep (\{J_1^{\lambda},J_2^{\lambda}\})\right).
\end{equation}
Now, we consider the asymptotic case   $\ep \rightarrow 0$. All
operators and equations (\ref{J0})-(\ref{qcom3}) in this limit are
represented as infinite matrices in the basis $\left\{|m\rangle
\right\}$ of the abstract Hilbert space $\mathcal{H}$. The equations
(\ref{qcom1})-(\ref{qcom2}) prove that in the asymptotic case the
mapping (\ref{dsoper}) is a homomorphism sending the $su(1,1)$ Lie
Poisson algebra into the Lie algebra generated by $\left\{
\hat{J}_0^{\lambda},\hat{J}_1^{\lambda},\hat{J}_2^{\lambda}\right\}$,
where $\hat{J}_a^{\lambda} :=
\lim_{\ep\rightarrow 0}\hat{J}^{\ep, \lambda}_a~~(a=0,1,2)$.

It has been proved in Ref. \onlinecite{GPW}, for $\lambda =0,$ that
the operators $\hat{J}_a^{\lambda}$ are essentially self-adjoint. The
corresponding proof for $\lambda =1/2$ can be done by analogy.

\subsection{``Angular'' representation}

In agreement with our general framework  we denote by
$\mathcal{K}_\ep$ the closure of the linear span of the orthonormal
set  $\left\{\phi_m^\ep \right\}_{m \in \dZ} $ given in
(\ref{phimep}).  The space $\mathcal{K}_\ep$ is a reproducing Hilbert
subspace of $L^2(X,\mu_\ep)$, isomorphic to $\mathcal{H}$ through the
one-to-one map
\begin{equation}
\label{mapkh} {\mathcal H} \ni | \phi \rangle  \rightarrow
{\phi^\ep}(\beta, J) \equiv \sqrt{\mathcal{N}_\ep(J)}\langle \beta,
J, \ep | \phi \rangle \in \mathcal{K}_\ep .
\end{equation}
The reproducing kernel for elements of $\mathcal{K}_\ep$ is given by
\begin{equation}
\label{reker} K_\ep( \beta, J;  \beta', J') = \sum_{m \in \dZ}  {
\phi_m^\ep} ( \beta, J) \overline{\phi_m^\ep ( \beta', J')} .
\end{equation}
If we specify the map (\ref{mapkh}) to the basis elements of
$\mathcal{H}$, we obtain
\begin{equation}
\label{baskh} | m \rangle  \rightarrow
\sqrt{\mathcal{N}_\ep(J)}\langle  \beta, J, \ep |m \rangle =
{\phi_m^\ep}(\beta, J) =
\big(\frac{\ep}{\pi}\big)^{1/4}\exp\big(-\frac{\ep}{2}(J-m)^2\big)
\exp (im\beta).
\end{equation}

It is clear from this correspondence that the  basis $| m\rangle$'s
could as well be considered as Fourier exponentials $e^{i m\beta}$
forming the orthonormal basis of the Hilbert space $L^2(\dS^1) \simeq
\mathcal{H}$. They are the \emph{spatial modes} in this ``angular
position'' representation and they are also obtained as the limit,
after suitable rescaling,
of the ${\phi_m^\ep}(\beta, J)$'s at $\ep = 0$.

In this representation, the operator $\hat{J}_0^{\ep, \lambda}$ is
nothing  but
the $\lambda$-shifted angular momentum operator: $\hat{J}_0^{\ep,
\lambda}= \lambda
-i\frac{\partial}{\partial \beta}$. On the other hand,
$A_\ep(e^{i\beta}) $ is multiplication operator by  $e^{i\beta}$ up
to the (arbitrarily close to 1) factor
$e^{-\frac{\epsilon}{4}}$.
     The ``canonical'' commutation rule
$$ \lbrack A_\ep(J), A_\ep(e^{i\beta}) \rbrack =
A_\ep(e^{i\beta})$$ is
canonical in the sense that it is in exact correspondence with the
classical Poisson bracket
$$\left\{ J, e^{i\beta} \right\} = i e^{i\beta}$$
\textit{It is actually the only non trivial one having this exact
correspondence.}

Although the function ``angle'' is strictly speaking not a classical
observable since it is not continuous at $\theta = 0 \, \mathrm{mod}
\, 2 \pi$, we take here the freedom to consider its quantum
counterpart defined by
     \begin{equation}
\label{opangle}
     A_\ep(\beta) = \pi \id_{{\mathcal H}} +  \sum_{m\neq m'}
i \frac{e^{-\frac{\epsilon}{4}(m-m')^2}}{m-m'}\, | m\rangle \langle m'
|.
     \end{equation}
Difficulties about correctly defining angle operator in quantum
mechanics and related questions of angular localization and
uncertainties are famous (see  Ref. \onlinecite{HAK}).
     As  a matter of fact, there could be serious interpretative
difficulties with commutation rules of the type
$$  \lbrack A_\ep(J), A_\ep(f(\beta)) \rbrack =  \sum_{m, m'}
     (m-m')\,e^{-\frac{\epsilon}{4}(m-m')^2} \,c_{m-m'}(f)\, |
m\rangle \langle m' |,$$
when $f$ is not as regular as a Fourier exponential.
In particular, we obtain for the angle operator:
\begin{equation}
\label{ccrcir}
     \lbrack A_\ep(J), A_\ep(\beta) \rbrack = i \sum_{m, m'}
     e^{-\frac{\epsilon}{4}(m-m')^2}\, | m\rangle \langle m' |,
     \end{equation}
to be compared with the classical $\left\{ J, \beta \right\} = 1$ !

Actually, these difficulties are just apparent and are due to the
discontinuity of the $2\pi$ periodic function $B(\beta)$ equal to
$\beta$ on $\lbrack 0, 2\pi [$. They can be circumvented if we
examine, for instance the behavior of lower symbols $\langle \beta',
   J' | A_{\beta} |  \beta', J' \rangle$ and $ \langle\beta',
   J' |  \lbrack A_J, A_{\beta} \rbrack | \beta',  J'
\rangle$ for $0 < \beta' < 2 \pi$ at the limit $\epsilon \to 0$. They
behave like $\beta'$ and $-i$, respectively.

\subsection{Group representation}

To identify the $SU(1,1)$ representation which integrates within the
Bargmann classification, we consider the Casimir operator. The
classical Casimir operator $C$ for the algebra (\ref{shifobs}) (as
well as for the algebra (\ref{clcom})) has the form
\begin{equation}\label{CC}
C=(J_2^{\lambda})^2 + (J_1^{\lambda})^2 - (J_0^{\lambda})^2 =\kappa ^2 .
\end{equation}
Making use of (\ref{dsoper}) we map $C$ into the corresponding  quantum
operator $\hat{C}$ as follows
\begin{eqnarray}\label{hCC}
\hat{C}:= \lim_{\ep\rightarrow 0}\hat{C}^\ep&:=&\lim_{\ep\rightarrow 0}
\left(\hat{J}_2^{\ep,\lambda} \hat{J}_2^{\ep,\lambda}
+\hat{J}_1^{\ep,\lambda} \hat{J}_1^{\ep,\lambda} -
\hat{J}_0^{\ep,\lambda} \hat{J}_0^{\ep,\lambda} \right)= \nonumber \\
& &\lim_{\ep\rightarrow 0}\sum_{m\in \dZ} \left(e^{-\ep /2}
(m +\lambda)^2 +\kappa ^2 +
\frac{1}{4})- (m +\lambda)^2 \right)|m><m|=\nonumber \\& & (\kappa ^2 +
\frac{1}{4})
\sum_{m\in \dZ} |m><m| =
(\kappa ^2 + \frac{1}{4})\id =: q \id .
\end{eqnarray}
Thus, we have obtained that our choice of coherent states
(\ref{cosit}) and the mapping (\ref{dsoper}) leads, as $\ep
\rightarrow 0$, to the representation of $su(1,1)$ algebra with the
Casimir operator to be an identity in $\mathcal{H}$ multiplied by a
real constant $1/4 < q < \infty$.

In the asymptotic case ($\ep \rightarrow 0 $) all operators are
defined in the Hilbert space $\mathcal{H}$. The specific realization
of $\mathcal{H}$ may be obtained by finding the space of
eigenfunctions of the set of all commuting observables of the system.
We choose $\hat{C}$ and $\hat{J}_0^{\lambda}$. It is easy to verify
that the
set of common eigenfunctions of $\hat{C}$ and $\hat{J}_0^{\lambda}$ is
independent of $\lambda$ and reads as
\begin{equation}\label{fmb}
\phi _{m}(\beta):= \frac{1}{2\pi}e^{i m\beta},
~~~m\in\dZ .
\end{equation}
The set of eigenfunctions (\ref{fmb}) is obviously complete
    in $\mathcal{H}$. Note that we could as well choose the pair
$\hat{C},~ \hat{J}_0 = - i\frac{d}{d\beta} = \hat{J}_0^{\lambda} -
\lambda$ with the basis
\begin{equation}\label{fmbshi}
\phi _{m}(\beta):= \frac{1}{2\pi}e^{i m\beta},
~~~m\in\dZ,
\end{equation}
which corresponds to different boundary conditions according to
whether $\lambda = 0$ or $\lambda = 1/2$. The Bargmann classification
\cite{VB} is characterized by the range of $q$ and $m$. Since $1/4 <
q < \infty$, the class with $m$ integer corresponds to the first
branch, $\lambda =0$, of the principal series  of $SO_0(1,2)$
group already presented in Section III. The second branch,
$\lambda =1$, is realized by the class with $m$ half-integer.

Comparing the results of quantization by group theoretical method
\cite{WP2} with the results of this section we can see that the
choice of the coherent states in the form induced by (\ref{cosit})
ensures that our coherent state results are time-reversal invariant.

\subsection{Vector coherent states}
In the previous sections, the occurrence of the parameter $\lambda$
looks rather artificial.
   Let us indicate here another way of defining the coherent
states which takes into account the time-reversal invariance in a more
natural fashion. It is
based on the vector coherent state construction (see Ref.
\onlinecite{AEG}):

Apart from a Hilbert space $\mathcal{H}$ with an orthonormal basis
$\{|m>\}_{m\in\dZ}$, we introduce the Hilbert space $\dC^2$ with the
orthonormal basis $\chi^k ~ (k=1,2)$ defined as
\begin{equation}\label{k12}
\chi^1 =  \left( \begin{array}{c} 1 \\0 \end{array} \right),~~~~~
\chi^2 =  \left( \begin{array}{c} 0 \\1 \end{array} \right)
\end{equation}
The set of vectors $\chi^k \otimes |m> ~~(k=1,2; \: m\in\dZ)~ $
forms an orthonormal basis of $\dC^2\otimes \mathcal{H}$.

Next, we define a Hilbert-Schmidt $F_m^\ep$ type operator on $\dC^2$
\begin{equation}\label{Xx}
X\ni x \longrightarrow F_m^\ep (x)\in \mathcal{B}(\dC^2),~~~~m\in\dZ ,
\end{equation}
    as follows
\begin{equation}\label{fmx}
F_m^\ep (x) := \overline{\phi^\ep_m (x)}\left(
\begin{array}{cc}1 & 0
\\ 0 & e^{i\beta /2}  \end{array} \right),
\end{equation}
    where $\phi^\ep_m $ are defined by (\ref{phimep}).
$\mathcal{B}(\dC^2)$ is the space of Hilbert-Schmidt operators on
$\dC^2$
with the scalar product
\begin{equation}\label{YZ}
<Y|Z> := Tr [Y^\ast Z],~~~~~Y,Z \in \mathcal{B}(\dC^2),
\end{equation}
where $Tr [Z]:= (\chi^1)^\ast Z \chi^1 +  (\chi^2)^\ast Z \chi^2$.

Then, we define $\widetilde{\mathcal{N}_\ep}$  corresponding to
(\ref{theta}) as
\begin{equation}\label{nte}
\widetilde{\mathcal{N}_\ep}(x) := \sum_{m\in \dZ}Tr[|F_m^\ep (x)|^2]=
2\mathcal{N}_\ep (x) <\infty ,
\end{equation}
where $|F_m^\ep (x)|:= [F_m^\ep (x)F_m^\ep (x)^\ast]^{1/2}$ denotes the
positive part of the operator $F_m(x)$.

\noindent
One may verify that
\begin{equation}\label{intX}
\int_X \mu(dx)F_m^\ep (x) F_n^\ep (x)^\ast =\delta_{mn}\id_{\dC^2},
~~~~~m,n \in \dZ ,
\end{equation}
where $\id_{\dC^2}$ denotes the identity operator in $\dC^2$.

Finally, the vector coherent states $|x,\ep ;\chi^k>\in
\dC^2\otimes \mathcal{H}$ are defined as (see Ref. \onlinecite{AEG}
for details)
\begin{equation}\label{x12}
|x,\ep ;\chi^k> :=\widetilde{\mathcal{N}_\ep}(x)^{-1/2}\sum_{m\in
\dZ} F_m^\ep (x)\; \chi^k \otimes |m>,~~~~~k=1,2.
\end{equation}
One may check that due to (\ref{nte}) and (\ref{intX}) the states
(\ref{x12}) are coherent, i.e. they satisfy the normalization
condition and lead to the resolution of the identity in
$\dC^2\otimes \mathcal{H}$ :

\begin{equation}
\label{VCSDS}
   \sum_{k=1}^{2}\langle x , \ep ;\chi^k\; \vert\; x,  \ep ;\chi^k\rangle
= 1, ~~~    \ \sum_{k=1}^{2}\int_X
\mu(dx)\widetilde{\mathcal{N}_\ep}(x)\;\vert x , \ep ;\chi^k
\rangle\langle x , \ep ;\chi^k \vert =
            \mathbb{I}_{\dC^2}\otimes\mathbb{I}_{\mathcal{H}}
\end{equation}

Extension of the method to the finite or infinite covering of
$SU(1,1)$ is straightforward. Now, the quantization of non-shifted
classical observables $J_0, J_1, J_2$ through the computation  of the
operator-valued integrals
\begin{equation}\label{VCSQuant}
\left\lbrace\begin{array}{c }
        \widetilde{J_0^\ep}    \\
        \widetilde{J_1^\ep} \\
        \widetilde{J_2^\ep}
\end{array} \right\rbrace
:=
   \sum_{k=1}^{2}\int_X  \mu(dx)
\widetilde{\mathcal{N}_\ep}(x)
\left\lbrace\begin{array}{c }
        J_0    \\
        J_1 \\
        J_2
\end{array} \right\rbrace
\;\vert x , \ep ;\chi^k \rangle\langle x , \ep
;\chi^k \vert
\end{equation}
leads to three operators acting on $\dC^2\otimes \mathcal{H}$. Direct
calculations yields:
\begin{equation}\label{vv0}
\widetilde{J_0^\ep}= \sum_{k=1}^{2}\sum_{m\in \dZ} m\; \chi^k \otimes
|m><m|\otimes \overline{\chi^k} ,
\end{equation}
\begin{equation}\label{vv1}
\widetilde{J_1^\ep}= \frac{1}{2}e^{-\ep/4}\sum_{k=1}^{2}\sum_{m\in
\dZ} \left( (m+1/2 + i\kappa)\; \chi^k \otimes |m+1><m|\otimes
\overline{\chi^k} + c.c. \right) ,
\end{equation}
\begin{equation}\label{vv12}
\widetilde{J_2^\ep}= \frac{1}{2i}e^{-\ep/4}\sum_{k=1}^{2}\sum_{m\in
\dZ} \left( (m+1/2 + i\kappa)\; \chi^k \otimes |m+1><m|\otimes
\overline{\chi^k} - c.c. \right) ,
\end{equation}
where $c.c.$ stands for the Hermitian conjugate of the preceding
term.

\noindent Hence,  both possibilities, $\lambda = 0$ and $\lambda =
1/2$, are considered within the same Hilbertian framework.

\section{Prequantization}

\subsection{Homomorphism and irreducibility}

For the choice of parametrization of the phase space $X$ in the form
(\ref{paraph}), there exists  well defined symplectic potential
$\theta$ and 2-form $\omega$ given by
\begin{equation}\label{the}
\theta:=Jd\beta,~~~\omega:=d\theta = dJ\wedge d\beta.
\end{equation}
The volume element  of the symplectic manifold $(X,\omega)$ is
normalized as
\begin{equation}\label{dmu}
\mu(dJd\beta):=\omega/2\pi .
\end{equation}
The Hilbert space  associated with $(X,\omega)$ is defined to be
$L^2(X,\mu)$ with the standard scalar product. The Hamiltonian vector
fields $X_a ~(a=0,1,2)$ corresponding to the observables
(\ref{param})  are solutions to the equations \cite{WNM}
\begin{equation}\label{Xa}
X_a\rfloor\omega +dJ_a =0,~~~~~a=0,1,2.
\end{equation}
They  are found to be
\begin{equation}\label{X12}
X_1=\cos\beta \:X_0 + (J_0
\sin\beta+\kappa\cos\beta)\:\frac{\partial}{\partial J},
~~~X_2=\sin\beta \:X_0-(J_0\cos\beta-\kappa\sin\beta)
\:\frac{\partial}{\partial J} ,
\end{equation}
and
\begin{equation}\label{X0}
X_0=\frac{\partial}{\partial\beta},~~~J_0 :=J+\lambda,~~~\lambda
=0,~1/2 .
\end{equation}
We choose $\lambda=0$ or $\lambda=1/2$ to fit the previous sections.
The pre-quantum observables $\tilde{J}_a$ corresponding to
(\ref{param}) are defined \cite{WNM} by
\begin{equation}\label{Jat}
\tilde{J}_a :=-i X_a-X_a\rfloor\theta + J_a,~~~~a=0,1,2
\end{equation}
and finally read
\begin{equation}\label{J0t}
\tilde{J}_0=-i \frac{\partial}{\partial\beta}+ \lambda,
\end{equation}
\begin{equation}\label{J1t}
\tilde{J}_1=\cos\beta \:\tilde{J}_0
-\kappa\sin\beta-i(J_0\sin\beta +\kappa
\cos\beta)\:\frac{\partial}{\partial J},
\end{equation}
\begin{equation}\label{J2t}
\tilde{J}_2=\sin\beta \:\tilde{J}_0 +\kappa\cos\beta
+i(J_0\cos\beta -\kappa \sin\beta)\:\frac{\partial}{\partial J} .
\end{equation}
It is known \cite{WNM} that (\ref{Jat}) leads to the homomorphism.
Thus  we have
\begin{equation}\label{tcom}
[\tilde{J}_a,\tilde{J}_b]= -i \widetilde{\{J_a,J_b\}},~~~~a,b=0,1,2.
\end{equation}
The quantum Casimir operator $\tilde{C}$ corresponding to prequantum
operators (\ref{J0t})-(\ref{J2t}) reads
\begin{equation}\label{tCas}
    \tilde{C}:=\tilde{J}_2^2 +\tilde{J}_1^2-\tilde{J}_0^2 =({\kappa}^{2}
+i\kappa)
-( {\kappa}^{2}+{J_0}^{2}  ) \:{\frac {\partial ^{2}}{
\partial {J}^{2}}} -2 \kappa \:{\frac {\partial ^{2}}
{\partial J\partial \beta}} +2(i\kappa J - J_0)\:{\frac {\partial
}{\partial J}}.
\end{equation}
Now we intend to find   a suitable dense subspace
$\widetilde{\mathcal{H}}$ of the Hilbert space $L^2(X,\mu)$ (if the
algebra consists of unbounded operators that is our case) such that
the lift of the (essentially) self-adjoint representation of the
algebra to the unitary representation of the corresponding Lie group
is \textit{irreducible}.  Here, we try to find   a dense subspace
$\widetilde{\mathcal{H}}\subset L^2(X,\mu)$ which consists of the
functions $G_{\lambda\kappa m}$ satisfying  the following two
equations
\begin{equation}\label{EK1}
\tilde{C}\:G_{\lambda\kappa m}(\beta,J) = (\kappa^2 +1/4)
\:G_{\lambda\kappa m}(\beta,J),
\end{equation}
\begin{equation}\label{EK2}
       \tilde{J_0}\:G_{\lambda\kappa m}(\beta,J)=m\:G_{\lambda\kappa
m}(\beta,J),
       ~~~~m\in \dZ.
\end{equation}

\noindent We impose (\ref{EK1}) and (\ref{EK2}) because we intend to
get the connection with the principal series representation of
$sl(2,\dR)$ algebra in the Bargmann form \cite{VB}. In the case where
$\{G_{\lambda\kappa m}\}_{m\in\dZ}$ consists of a total set of
orthonormal functions, it can be used as a basis of
$\widetilde{\mathcal{H}}$.

\subsection{Square-integrability}

Now,  we examine  the square-integrability of  the functions
$G_{\lambda\kappa m}$.

\noindent First, let us look for the  non-trivial common solutions to
(\ref{EK1}) and (\ref{EK2}), in the form:
\begin{equation}\label{E5}
G_{\lambda \kappa m}(\beta , J)=e^{i(m-\lambda)\beta} \:f_{\lambda
\kappa m} (J) .
\end{equation}
    Replacing $~\partial /\partial \beta~$ in (\ref{EK1}) by
$~\partial /\partial \beta~$ of (\ref{EK2}),  using (\ref{E5}) and
making simple algebraic rearrangement of the terms in the resulting
form of (\ref{EK1}), we finally obtain the equation
\begin{equation}\label{E6}
       (J-\sigma)(J-\overline{\sigma})\:\frac{\partial^2 f}{\partial
       J^2}+ (\delta + \rho J)\:\frac{\partial f}{\partial J}+ \tau f =0,
\end{equation}
where
\[ \sigma := -\lambda - i \kappa,~~~\overline{\sigma} = -\lambda +i
\kappa,~~~
\delta := 2\lambda + 2i\kappa(m-\lambda),~~~\rho :=
2(1-i\kappa),~~\tau~:=1/4 -i\kappa ,\] and where $~f(J):=f_{\lambda
\kappa m}(J)~$ for fixed values of $~\lambda ,~\kappa ~$ and $~m$.

It is clear that (\ref{E6})
is the Gauss equation with the three regular singularities, namely at
$J=\sigma,\; \overline{\sigma}\;$ and $\;\infty\;$. Its solution may
be obtained  from the solution to the standard Gauss equation
\begin{equation}\label{E7}
       z(1-z)\:\frac{d^2
       h(z)}{dz^2}+\big(c-(a+b+1)z \big)\:\frac{dh(z)}{dz}-ab\:h(z)
       =0,
\end{equation}
having three regular singularities at $z=0,\;1\;$ and $\;\infty$. For
this purpose  we first rewrite (\ref{E7}) in the form of (\ref{E6}),
and then compare the coefficients of the corresponding derivatives to
get the expressions for $\:a,\:b\:$ and $\:c\:$:

\noindent Making use of the transformation
\begin{equation}\label{S2}
       \dC\ni z\longrightarrow J(z):= \sigma +2i\kappa z,
\end{equation}
we map the singular points $(\:0,\:1,\:\infty\:)$ of (\ref{E7}) onto
the singular points $(\:\sigma,\:\overline{\sigma},\:\infty\:)$ of
(\ref{E6}).  The equation (\ref{E7}) in terms of the variable
$\:J\:$, defined by (\ref{S2}),  reads
\begin{equation}\label{S3}
       (J-\sigma)(J-\overline{\sigma})\:\frac{\partial^2 f}{\partial
       J^2}+ (-2i\kappa c +(1+a+b)(J-\sigma))\:\frac{\partial f}
       {\partial J}+ ab\: f =0,
\end{equation}
where
\[f(J):=h\big(\frac{J-\sigma}{2i\kappa}\big).\]
    The comparison of (\ref{E6}) and (\ref{S3}) leads to
\begin{equation}\label{S4}
       a=1/2 - 2\kappa i
       ,~~~b=1/2,~~~c=1-m-\kappa i .
\end{equation}
Finally, the solution of (\ref{E6})  may be obtained by the insertion
of $\:z=(J-\sigma)/2i\kappa\:$  and $(\:a,\:b,\:c\:)$ defined by
(\ref{S4}) into  the solution of (\ref{E7}).

As it is known \cite{ABC,AFN},  in case none of the numbers
$~a,~b,~c-a, ~c-b~$ is an integer (our case) the two linearly
independent solutions of (\ref{E7}) can be obtained from any
non-trivial solution by analytic continuation along a path which
encircles one of the point $z=0,1,\infty$. This way one can get the
solution of (\ref{E7}) in the neighborhood of $\:z=0\:$ in the form
\begin{equation}\label{E8}
       h_1(z)=A_1\:F(a,b,c;z)+B_1\:z^{1-c}F(a+1-c,b+1-c,2-c;z),
\end{equation}
with the convergence range $\:|z|<1\:$.
    In the neighborhood of $\:z=1\:$ the solution reads
\begin{equation}\label{E9}
h_2(z)=A_2\:F(a,b,1+a+b-c;1-z)+B_2\:(1-z)^{c-a-b}F(c-b,c-a,1+c-a-b;1-
z),
\end{equation}
with  $\:|z-1|<1\:$.
    Finally, in the neighborhood of $\:z=\infty\:$ the solution is the
following
\begin{equation}\label{S0}
h_3(z)=A_3\:z^{-a}\:F(a,1+a-c,1+a-b;z^{-1})+B_3\:z^{-b}\:F(b,1+b-c,1+b-
a;z^{-1}),
\end{equation}
with  $\:|z|>1\:$.

\noindent The coefficients $\:A_k,\:B_k~~(k=1,2,3)~~$ in
(\ref{E8})-(\ref{S0}) are any complex numbers, whereas
$\:F(a,b,c;z)\:$ is the well known hypergeometric function defined as
\begin{eqnarray} \label{S1}
       F(a,b,c;z)& = &
       1+\frac{ab}{1!\:c}\:z+\frac{a(a+1)b(b+1)}{2!\:c(c+1)}\:z^2+\ldots
       \\ \nonumber
       &+& \frac{a(a+1)\cdots (a+n-1)b(b+1)\cdots
(b+n-1)}{n!\:c(c+1)\cdots
       (c+n-1)}\:z^n+\ldots.
\end{eqnarray}

At this stage we are ready to discuss the square-integrability of
$G_{\lambda \kappa m}$. For that purpose we  consider the following
integral
\begin{equation}\label{S5}
I:= \frac{1}{2\pi}\:\int_0^{2\pi}d\beta\:\int_{-\infty}^\infty
dJ|G_{\lambda \kappa m}(\beta ,J)|^2 = \int_{-\infty}^\infty dJ
|f_{\lambda\kappa m}(J)|^2 .
\end{equation}
We do not intend to calculate $~I~$ but only  make estimate of it. In
the neighborhood of $~z=\infty~$ the solution (\ref{S0}, due to
(\ref{S1}), reads
\begin{equation}\label{S7}
h_3(z)= A_3\:z^{-a}+B_3\:z^{-b} + \mathcal{O}(z^{-a-1}) +
\mathcal{O}(z^{-b-1}) .
\end{equation}
For large value of $|J|$  the function $~f_{\lambda\kappa m}~$ (with
$\:a,b,c\:$ defined by (\ref{S4}) and $\:z=(J-\sigma)/2i\kappa\:)$,
due to (\ref{S7}), has the property
\begin{equation}\label{S6}
\left\vert f_{\lambda\kappa m}\left(\frac{J-\sigma}{2\pi
i}\right)\right\vert^2 =
\mbox{const}\cdot |J|^{-1} + \mathcal{O}(|J|^{-2 }).
\end{equation}

\noindent  Since the singular points $~\sigma~$ and
$~\overline{\sigma}~$ do not belong to the real axis of $J,$ the
integrand of (\ref{S5})  is a well defined  analytic function for
finite $J$. Therefore, due to (\ref{S6}), the integral (\ref{S5})
diverges logarithmically.  It means that the non-trivial solutions
$~G_{\lambda \kappa m}~$ to (\ref{EK1}) and (\ref{EK2}) are not
square-integrable with respect to the standard measure (\ref{dmu}).

\section{Merits and demerits of quantization schemes}
The \textit{group theoretical method} is  powerful, and currently
used by physicists. It operates when  the search of the
representation of the algebra of observables of a physical system is
the main concern, and  when an  irreducible (possibly projective)
unitary representation of the corresponding Lie group is available in
the mathematical literature. In such a case,  the Stone theorem leads
to an (essentially) self-adjoint representation of the algebra. Our
Sec. III    demonstrates the efficiency of such a method. However, it
is not always so simple.  Although  physicists usually identify the
Lie algebra of observables,   the representation of the corresponding
Lie group may remain  unknown. Difficulties occur when the  classical
observables are higher than second  order polynomials in  the
canonical variables \cite{VHL}. In this case,   a problem appears
with the ordering of the canonical operators  mapping the classical
observables into the corresponding quantum operators (see, e.g. Ref.
\onlinecite{EEM} and references therein). Then, a general rule is to
choose the ordering which provides  self-adjoint operators in a
certain dense subspace of a Hilbert space (quantum observables are
usually unbounded operators). However, the proof of self-adjointness
may happen to be quite involved. It may require examination of
solutions to the deficiency indices equations, which for some
orderings and non trivial functional forms of observables are very
complicated non-linear equations. Clearly,   this may provide  more
than one representation of the algebra of observables and it is
sometimes difficult to prove their unitary (non)equivalence. In
principle an ambiguity may be lifted by comparison with experimental
data, but in many cases one considers only toy models to `get insight
into the physical problem'. In our case, for the  particle dynamics
on hyperboloid, this problem was solvable because we found   two
canonical variables,  such that the observables are linear functions
in one of them. Making use of the Schr\"{o}dinger representation for
the canonical variables and the symmetrization prescription for
operators ordering we have obtained the deficiency indices equations
easy to solve \cite{WP1}.

The \textit{coherent state quantization method} presents  no problem
with the ordering of operators, since the mapping (\ref{oper})
includes only classical expression for observables. The main
difficulty  is the definition of the coherent states, in particular
the construction of the set of orthonormal vectors which spans a
closed subspace ${\cal H}$ of the Hilbert space $L^2(X,\mu)$. Here,
we have solved this problem, following  inspiration by   the choice
of coherent states for dynamics of a particle on a circle.
Application of coherent state quantization for a particle on a sphere
may be found in Refs. \onlinecite{KR2,BCH}. For general structure of
the coherent state quantization method  we recommend Refs.
\onlinecite{A3,berez,A1,A2,GGH,AEG}.

The \textit{geometric quantization method} (and to some extent the
coadjoint orbit method in the Lie group theoretical context
\cite{BBK,AAK,JMS}) has the ambition of being a rigorous canonical
quantization method. Its main problem is that at prequantization step
it leads to a \emph{reducible} representation of the symmetry group.
One cannot remove this reducibility in any nontrivial physical
situations  \cite{JHD,WNM,GMT}. In this paper we have demonstrated
that the standard method leads to reducibility in the considered
case. By standard method,  we mean the determination of quantum
operators by prequantization rule and the choice of a measure
identified to the symplectic two-form.  It can be argued that another
measure,  or another mapping of basic observables, would do better.
But different choices for the  quantum observables, or for the
measure defining the scalar product, could easily spoil the
self-adjointness of quantum operators and the homomorphism of the
prequantization mapping.  The geometric quantization procedure
suffers from another demerit: it applies comparatively easily only
when canonical variables exist which are well defined globally. We
emphasize that the coherent state method does not require global, or
canonical variables on the phase space. However, it does not
guarantee  that the mapping of classical observables into
corresponding operators is always a homomorphism (in case considered
in this paper the mapping is the homomorphism). The geometric
quantization method always leads, by its construction, to the
homomorphism.

For other discussions concerning comparison  of the three methods
considered in our paper we recommend Ref. \onlinecite{EEM} (general
case) and Ref. \onlinecite{HAK} (case of $sl(2,\dR)$ algebra).

\begin{acknowledgments}
W.P. would like to thank  the APC of the University of Paris VII  for
its hospitality and  the CNRS for financial support. He is also
pleased to acknowledge helpful discussions with S. T. Ali, J.
Derezi\'{n}ski, E. Huguet, J. Rembieli\'{n}ski and  J. Renaud.

\end{acknowledgments}

\end{document}